\documentclass[useAMS,usenatbib]{mn2e}
\usepackage{graphicx}
\usepackage{float}
\usepackage[usenames,dvipsnames]{xcolor}

\title[1H0707-495 with {\em NuSTAR}]{The Compton hump and variable blue wing in the extreme low-flux {\em NuSTAR} observations of 1H0707-495}
\author[Kara et al.]{E. Kara$^{1}$\thanks{E-mail:
ekara@ast.cam.ac.uk}, A. C. Fabian$^{1}$, A. M. Lohfink$^{1}$, M. L. Parker$^{1}$, D. J. Walton$^{2}$,
\newauthor
S. E. Boggs$^{3}$, F. E. Christensen$^{4}$, C. J. Hailey$^{5}$, F. A. Harrison$^{2}$, G. Matt$^{6}$,
\newauthor
C. S. Reynolds$^{7,8}$, D. Stern$^{9}$, and W. W. Zhang$^{10}$ \\
$^{1}$Institute of Astronomy, The University of Cambridge, Madingley Road, Cambridge, CB3 OHA\\
$^{2}$Cahill Center for Astronomy and Astrophysics, California Institute of Technology, Pasadena, CA 91125, USA\\
$^{3}$Space Science Laboratory, University of California, Berkeley, CA 94720, USA\\
$^{4}$DTU Space National Space Institute, Technical University of Denmark, Elektrovej 327, DK-2800 Lyngby, Denmark\\
$^{5}$Columbia Astrophysics Laboratory, Columbia University, New York, NY 10027, USA\\
$^{6}$Dipartimento di Matematica e Fisica, Universit\`a degli Studi Roma Tre, via della Vasca Navale 84, 00146 Roma, Italy\\
$^{7}$Department of Astronomy, University of Maryland, College Park, MD 20742-2421, USA\\
$^{8}$Joint Space-Science Institute (JSI), College Park, MD 20742-2421, USA\\
$^{9}$Jet Propulsion Laboratory, California Institute of Technology, Pasadena, CA 91109, USA\\
$^{10}$NASA Goddard Space Flight Center, Greenbelt, MD 20771, USA\\
}

\begin{document}

\date{\today}

\pagerange{\pageref{firstpage}--\pageref{lastpage}} \pubyear{2014}

\maketitle

\label{firstpage}

\begin{abstract}
The Narrow-line Seyfert I galaxy, 1H0707-495, has been well observed in the 0.3--10~keV band, revealing a dramatic drop in flux in the iron~K$\alpha$ band, a strong soft excess, and short timescale reverberation lags associated with these spectral features.  In this paper, we present the first results of a deep 250~ks {\em NuSTAR} observation of 1H0707-495, which includes the first sensitive observations above 10~keV.  Even though the {\em NuSTAR} observations caught the source in an extreme low-flux state, the Compton hump is still significantly detected.  {\em NuSTAR}, with its high effective area above 7~keV, clearly detects the drop in flux in the iron K$\alpha$ band, and by comparing these observations with archival {\em XMM-Newton} observations, we find that the energy of this drop increases with increasing flux.  We discuss possible explanations for this, the most likely of which is that the drop in flux is the blue wing of the relativistically broadened iron~K$\alpha$ emission line.  When the flux is low, the coronal source height is low, thus enhancing the most gravitationally redshifted emission.

\end{abstract}


\section{Introduction}
\label{intro}




Since the discovery of the broad iron~K line twenty years ago \citep{tanaka95}, there has been debate as to the origin of this distinctive spectral feature. Two physical interpretations fit the time-integrated energy spectrum equally well: relativistic reflection, where the iron~K$\alpha$ emission line is produced in the inner accretion disc and is asymmetrically broadened due to special and general relativistic effects in the strong potential well of the central supermassive black hole \citep[e.g. ][]{fabian89,brenneman06,dauser12}, and partial covering absorption, where the primary emission is partially covered by distant clumpy clouds or outflowing winds, and this absorption pattern mimics the shape of the broad iron line \citep[e.g.][]{inoue03,miller08}.

This debate comes to a head in the spectral fitting of Narrow-line Seyfert I galaxy, 1H0707-495 ($z=0.0411$). The time-integrated spectrum shows a dramatic drop in flux at 7~keV \citep{boller02}, which suggests in both the relativistic reflection model \citep{fabian04} and the partial covering model \citep{tanaka04} that there is an overabundance of iron in the emitting/absorbing material \citep[unless the absorbing material is outflowing with velocity $>0.3c$; ][]{done07}.  In the case of relativistic reflection, overabundant iron leads to a strong prediction: the broadened iron~L emission line at $\sim 0.9$~keV.  \citet{fabian09} showed in a 500~ks {\em XMM-Newton} that the residuals to a phenomenological continuum model show both the broadened iron~K and iron~L emission lines.  In response, \citet{mizumoto14} claim that these data can be well-fit by a partial covering by a Compton thin absorber that envelops a Compton thick absorber, however the major features in the spectrum (namely the iron-L and iron-K features) are fitted by ad-hoc absorption edges.  They say: ``In addition to the cold/thick and hot/thin absorbers that respectively explain the iron K- and L- edges, we had to introduce the {\em ad-hoc} edge components to explain the extremely strong edge features.'' \citet{mizumoto14} do not discuss the implications of adding the edges on the rest of the spectral model, nor do they discuss why the iron-L and iron-K edge energies vary independently if they are part of the same absorbing structure. 
The high-resolution {\em XMM-Newton}-RGS data were also examined, and \citet{blustin09} found no evidence for the narrow emission or absorption lines expected from a partial covering model\footnote{\citet{mizumoto14} also examined the same RGS dataset, and claim that there {\em are} narrow absorption features present. However, these authors do not include a statistical comparison between a featureless continuum and a model with absorption lines, so the significance of the putative narrow features cannot be accessed.}.

Fortunately, 1H0707-495, in addition to its striking spectral features, is also one of the most highly variable Seyfert galaxies, commonly known to increase by an order of magnitude in flux on timescales as short as one hour.  This allows for extensive spectral-timing analysis, which can help break the degeneracies of traditional spectral modelling.  In addition to revealing the iron~L feature in the soft excess, \citet{fabian09} and \citet{zoghbi10} showed that the high-frequency variations in the soft excess from 0.3--1~keV lag behind those in the 1--4~keV band by $\sim 30$~s.  These authors interpreted this short timescale lag as the reverberation time delay between the primary-emitting corona and the inner accretion disc.  This short time delay would put the corona at a height of $<10 r_{\mathrm{g}}$ from the accretion disc  \citep[using the black hole mass estimate of $2 \times 10^{6} M_{\odot}$ from the $M_{\mathrm{BH}}$--X-ray excess-variance relation; ][]{bian03}.  Later, as predicted by the relativistic reflection model, a short timescale lag was also found in the iron~K band using a total of 1.3~Ms of archival {\em XMM-Newton} data on the source \citep{kara13a}.  This timing analysis showed that the high-frequency variability in the soft excess is correlated with that of the iron~K band, and that both of these bands lag behind the 1--3~keV band by tens of seconds.  In other words, the energy spectrum of the high-frequency lags resembles the relativistic reflection energy spectrum. To date, the simultaneous soft excess and iron~K lag has not been explained with a partial covering model.

\citet{miller10} and \citep[and later][for a similar source Ark~564]{legg12} presented a partial covering timing model in which longer, low-frequency lags were interpreted as reverberation between a compact continuum source and distant circumnuclear material. In this model, the amplitude of the low-frequency lag increases with energy because the continuum photons travel through a medium whose opacity decreases with increasing energy, and therefore the fraction of scattered photons increases with energy.  We note that this partial covering model requires the corona to be compact because the high-frequency variability must be intrinsic to the primary emission.  In this model, the high-frequency time lags are simply a phase-wrapping artefact from the low-frequency lags.  If this is the case, then the high-frequency lags would be a mirror-image of the low-frequency lags (just with a smaller amplitude), however, as discussed in the previous paragraph, the high-frequency lags show a distinct signature in the iron~K band (for more detail, see \citealt{zoghbi11}; \citealt{kara13c}, and a complimentary argument based on lag measurements of NGC~6814 in \citealt{walton13b}). The high-frequency iron~K lag has now been well-established in a number of source (e.g. \citealt{zoghbi12,kara13b,zoghbi13,alston14}, and see \citealt{uttley14} for a review). {\em Until partial covering can explain both the high- and low-frequency spectral-timing results that have now been widely observed, it cannot be considered a viable physical model.} 

\begin{figure*}
\includegraphics[width=0.75\textwidth]{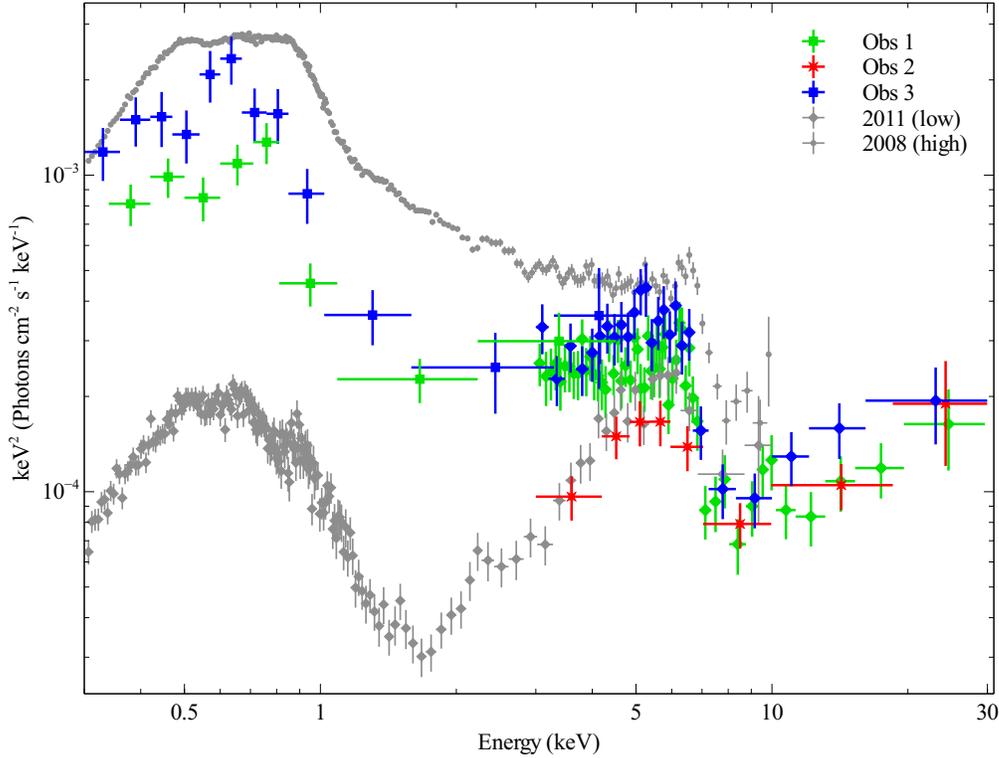}
\caption{Unfolded spectrum of 1H0707-495 to a powerlaw model with index 0 and normalization 1.  The green, red and blue points show Obs 1, Obs 2 and Obs 3 of our joint {\em NuSTAR} (combined FPMA and FPMB) and {\em Swift} observations.  The grey points show two archival {\em XMM-Newton} observations of one of the highest and lowest flux in order to demonstrate the range of fluxes exhibited by this source to date. }
\label{1h_nustar}
\end{figure*}

\begin{table*}
\centering
\begin{tabular}{c c c c c}
\hline
{\bf Name} & {\bf Date} & {\bf obsid} & {\bf Exposure (ks)} & {\bf FPMA/FPMB count rate (counts/s)}\\
\hline
Obs 1 & 2014 May 5 & 60001102002 & 144 & $0.008 \pm 0.0003/0.007 \pm 0.0003$\\
Obs 2 & 2014 June 10 & 60001102004 & 49 & $0.005 \pm 0.0004/0.005 \pm 0.0004$\\
Obs 3 & 2014 June 28 & 60001102006 & 47 & $0.01 \pm 0.0005/0.009 \pm 0.0005$\\
\hline
\end{tabular}
\caption{Details on the {\em NuSTAR} observations of 1H0707-495. Obs~1 and Obs~3 were accompanied by {\em Swift}-XRT snapshots.}
\label{obs}
\end{table*}

\citet{gardner14} recently presented an alteration to the reverberation lags in 1H0707-495, in which reverberation off the accretion flow around a non-spinning black hole is diluted by an occulting failed wind at $20~r_{\mathrm{g}}$. This model does require inner disc reflection, but the reflection fraction is very low because there is no emission from within $6 r_{\mathrm{g}}$ or more.  The authors do not explain the strong, very broad iron~K lag that is present in this source and many others.

Despite the extraordinary spectral features and variability patterns of 1H0707-495, it is a relatively faint source with a very steep spectrum, which means that it has never been detected above 10~keV until now \citep[e.g. ][]{walton13}.  The {\em Nuclear Spectroscopic Telescope Array} or {\em NuSTAR} \citep{harrison13} has the ability to make this detection because it carries the first high-energy focusing X-ray telescope in orbit, and is 100 times more sensitive than previous instruments in the 10--80~keV band.  Above 10~keV, photons are not photoelectrically absorbed and reprocessed. Rather they Compton downscatter, which produces an excess of photons in the 10--80~keV range.  We call this hard excess the Compton `hump', and it has been confirmed by {\em NuSTAR} in a number of objects, including NGC~1365 \citep{risaliti13,walton14}, MCG-6-30-15 \citep{marinucci14a}, and Mrk~335 \citep{parker14}.  {\em NuSTAR} even detects the associated high-frequency Compton hump reverberation lag in the bright sources, MCG-5-23-16 \citep{zoghbi14} and SWIFT~J2127.4+5654 \citep{kara14}, which confirms that the variability in the iron~K and Compton hump bands are correlated, and that they lag behind the continuum by short time delays. This time delay corresponds to light travel distances of $<10 r_{\mathrm{g}}$.    

We present the first results of the {\em NuSTAR} observation of 1H0707-495, which clearly reveals the sharp drop above 6.4~keV and Compton hump.
Given the recent evidence from {\em NuSTAR} observations and reverberation measurements that support the relativistic reflection interpretation, we will focus on that model for the remainder of the paper.
We fit the spectrum with a relativistic reflection model in order to explore the consequences of including the high-energy emission and to check that the model works.  We do not make attempts to fit the spectrum with a partial covering model, as it is not the aim of this paper to determine which model is correct based on the {\em NuSTAR} spectrum alone.  

The paper is outlined as followed: in Section~\ref{obs_sec} we describe the new observations, and we present the spectral results in Section~\ref{results}.  We find that the energy of the blue wing of the iron line is dependent on flux in this source, and discuss possible reasons and implications for this observation in Section~\ref{discuss}.  Finally, we summarize our conclusions in Section~\ref{conclude}.

\section{Observations and Data Analysis}
\label{obs_sec}

1H0707-495 was observed with the {\em NuSTAR} Focal Plane Module A (FPMA) and Focal Plane Module B (FPMB) on three separate occasions for a total exposure of $\sim 250$~ks.  See Table~\ref{obs} for dates, obsids, exposures and background-subtracted count rates. The Level 1 data products were processed with the {\em NuSTAR} Data Analysis Software ({\sc nustardas} v.1.4.1), and the cleaned Level 2 event files were produced and calibrated with the standard filtering criteria using the {\sc nupipeline} task and CALDB version 20140715.  We chose a circular source region of radius 40 arcsec and a background region of radius 85 arcsec. The regions were the same for both detectors.  While there was spectral variability between the three observations, there is little spectral variability within each observation.  The three pairs of spectra were binned in order to oversample the instrumental resolution by at least a factor of 2.5 and to have a signal-to-noise ratio of greater than $3 \sigma$ in each spectral bin.   

\begin{figure}
\includegraphics[width=\columnwidth]{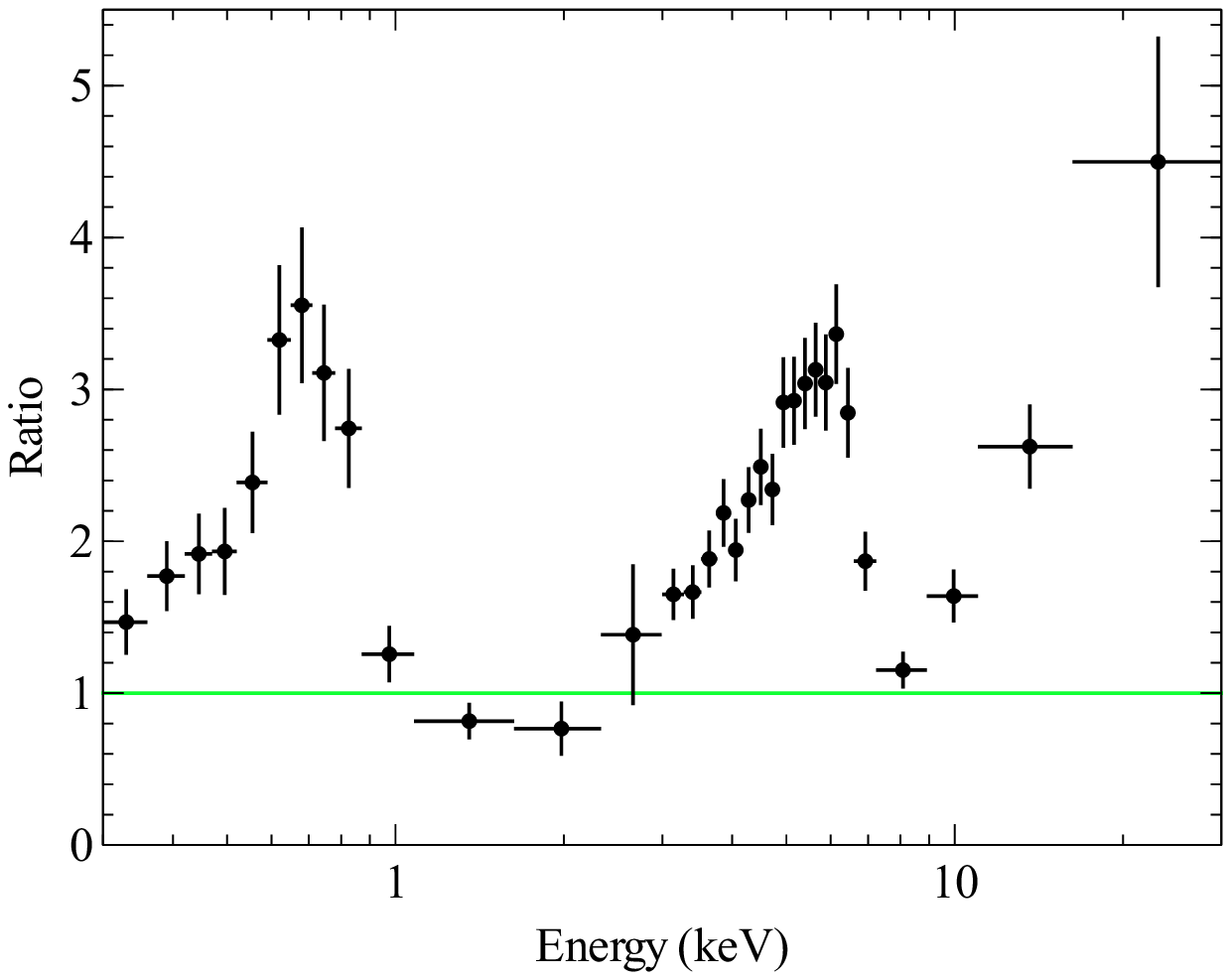}
\caption{The residuals to a phenomenological continuum model, revealing strong features at the iron~L, iron~K and Compton hump bands.  The data are from the {\em Swift} and {\em NuSTAR} observations of Obs~1 and Obs~3, which were taken in a similar flux state.}
\label{rat}
\end{figure}

{\em Swift}/XRT took simultaneous snapshots during Obs 1 (obsid 00080720001 and 00080720002 for 2~ks and 1.1~ks respectively) and Obs 3 (obsid 00080720004, for 1.7~ks). There was also a scheduled snapshot during Obs 2; however, this occurred during an anomaly period for {\em Swift}, and therefore the data could not be used for spectral analysis.  The {\em Swift} snapshots were processed using the automated processor supplied by the UK {\em Swift} Science Data Centre at the University of Leicester \citep{evans09}. 

We compare these new observations with archival {\em XMM-Newton} observations \citep{jansen01,struder01}.  1H0707-495 has been observed for a total of $\sim 1.5$~Msec with {\em XMM-Newton} over 15 separate observations. The data reduction of these archival observations is presented in \citet{kara13a}.  All error ranges are quoted at the 90 per cent confidence level for each parameter ($\Delta \chi^{2}=2.71$).

\section{Results}
\label{results}

All three observations of 1H0707-495 found the source in a low-flux state.  Fig.~\ref{1h_nustar} shows the unfolded spectra for our three {\em NuSTAR} observations along with the corresponding {\em Swift} snapshots (where possible).  Overplotted in grey are two archival {\em XMM-Newton} observations showing one of the highest (obsid: 0511580201) and lowest flux states (obsid: 0554710801) observed with {\em XMM-Newton}.   The second of the {\em NuSTAR} observations was the lowest flux observation of the three, and appears to be even lower than the previous lowest flux observation taken in 2011 as an {\em XMM-Newton} Target of Opportunity.  Unfortunately, we do not have the simultaneous {\em Swift} snapshot to confirm that Obs~2 was in a similar (or even lower) flux state at soft X-rays.  

{\em NuSTAR}, with its high effective area above 6~keV, clearly detects a strong drop in the emission around 6-7~keV, even when the source is in a low-flux state.  Interestingly, the energy of this drop is not constant between observations.  In the high-flux {\em XMM-Newton} observation, the drop occurs at a higher energy than in the low-flux observations.  This variable blue wing was first observed in early {\em XMM-Newton} observations \citep{gallo04}, and we confirm this difference now with additional high quality observations.  We interpret this as a variable blue wing of the iron~K$\alpha$ line, and discuss it further in Section~\ref{discuss}.

In Fig.~\ref{rat} we show the ratio between the combined Obs 1 and Obs 3 spectrum and the best fit continuum model, demonstrating the strong iron~L, iron~K and Compton hump features.  We did not include the Obs 2 {\em NuSTAR} spectrum because there is no accompanying {\em Swift} data.  The continuum model was produced by fitting the entire spectrum with a black body, powerlaw, two {\sc laor} models at iron~K and iron~L and a Gaussian for the Compton Hump.  The {\sc laor} lines and Gaussian were then removed, leaving just the continuum, consisting of a black body at kT=0.05~keV and powerlaw with $\Gamma=2.9$ with Galactic absorption.  Other sources observed with {\em NuSTAR} have shown similar ratio plots with clear iron~K and Compton hump features \citep[e.g.][]{risaliti13,walton14,parker14,marinucci14a,marinucci14b,balokovic14}.  The reflection features in 1H0707-495 are very pronounced ($\sim 4$ times that of the continuum).

Next, we fit the spectrum with a relativistic reflection model in {\sc XSPEC}.  The spectrum can be well described using the publicly available {\sc relxilllp}, which combines the xillver reflection grid \citep{garcia13} with the relconv\_lp relativistic convolution kernel for a point source corona above an accretion disc \citep{dauser13}.  The benefit of the lamppost model is that it returns a physical height to the corona and calculates the emissivity profile consistently from the source height, rather than assuming a powerlaw or broken powerlaw emissivity profile.  We emphasize that the lamppost geometry is an extreme simplification of the corona, which is likely to be compact, but obviously not a point source.  We allow for the reflection fraction (ratio of the reflected to primary emission, $F_r/F_p$) to vary because the self consistent reflection fraction from a point source corona will always over predict the true reflection fraction from a slightly extended corona.  

\begin{figure}
\includegraphics[width=\columnwidth]{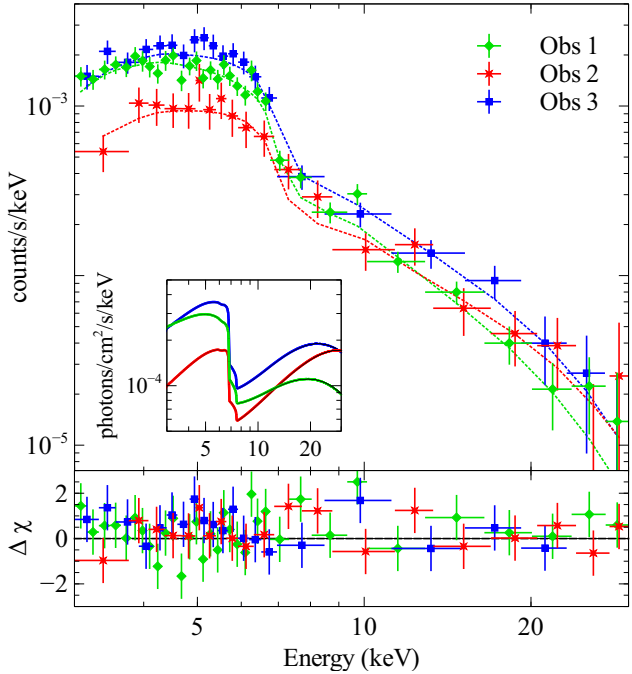}
\caption{The best fit relativistic reflection model, {\sc relxilllp} to the three {\em NuSTAR} observations.  The top panel shows the spectrum in counts, and the inset shows the unfolded model spectrum.  The bottom panel shows the $\chi^2$ residuals.  See Table~\ref{relxill} for more details on the fit.}
\label{mo}
\end{figure}

We start by fitting the three {\em NuSTAR} observations individually with the {\sc relxilllp} model, tying together physical parameters, such as the spin, inclination, and iron abundance. The goal was to see which parameters were changing between our three observations. Unfortunately, due to the quality of the data, we cannot constrain all parameters individually.  Instead, we start by fitting all three spectra simultaneously, tying all parameters together except the normalization.  This results in a good fit $\chi^2/\mathrm{dof}=251/233=1.08$. We use this as our baseline model, and test untying individual parameters that could change between our observations to see if this improves the fit.  The fit improves by letting the ionization parameter vary between observations.  This results in a $\Delta \chi^2=10$ for two additional parameters (i.e. the fit is preferred with $>99\%$ significance).  The ionization parameter is constrained to be lower for Obs 1 and Obs 3 than in Obs 2: log($\xi_{1})<0.9$ erg cm s$^{-1}$ and log($\xi_{3})=0.4\pm0.3$ for Obs~1 and Obs~3 versus log($\xi_{3})=1.75^{+1.25}_{-0.75}$ for Obs~2. The best overall fit comes from freeing the photon index $\Gamma$ between the different observations ($\chi^2=234/231$).  The improvement is $\Delta \chi^{2}=17$ for two additional parameters.    $\Gamma_{1}$ and $\Gamma_{3}$ are consistent within error, and $\Gamma_2$ is significantly harder, which is consistent with the observed flux and photon index correlation \citep{wilkins14}. See Fig.~\ref{mo} and Table~\ref{relxill} for the best fit {\sc relxilllp} model with normalization and photon index left free to vary between observations.  When we let only the source height vary between observations, the fit was slightly improved from our baseline model, but we could not find differences in the source height, as all were constrained to be less than $3 r_{\mathrm{g}}$ within the 90\% confidence limit.  

\begin{table}
\centering
\begin{tabular}{l l}
\hline
{\bf Parameter} & {\bf Value}\\
\hline
$h$ ($r_{\mathrm{g}}$) & $2.14^{+0.31}_{-0.09}$\\
$a$ &$0.97 \pm 0.03$ \\
$i$ ($^{\circ}$) & $43.4_{-3.9}^{+1.8}$\\
$r_{\mathrm{out}}$ $(r_{\mathrm{g}})^{\alpha}$& 400.0 \\
$\Gamma_{1}$ & $3.05^{+0.25}_{-0.1}$ \\
$\Gamma_{2}$ & $2.37^{+0.28}_{-0.17}$ \\
$\Gamma_{3}$ & $2.80^{+0.2}_{-0.05}$ \\
log($\xi$) (log(erg cm s$^{-1}$)) & $0.3^{+0.3}_{-0.2}$\\
$A_{\mathrm{Fe}}$ & $>7.2$ \\
$E_{\mathrm{cut}}$ (keV)$^{\alpha}$ & 300.0  \\
$R$ & $>5.8$\\
$N_{1} \times 10^{-4}$ &$ 3.6 \pm 1.4$\\
$N_{2} \times 10^{-4}$ &$ 0.52^{+0.53}_{-0.18}$\\
$N_{3} \times 10^{-4}$ &$ 2.0^{+1.6}_{-0.2}$\\
\hline
$\chi^{2}/\mathrm{dof}$ & 234/231 = 1.02 \\
\hline
\multicolumn{2}{l}{$^{\alpha}$ frozen}\\
\end{tabular}
\caption[Best-fit Spectral Parameters]{Best-fit spectral parameters of {\sc relxilllp} model, corresponding to Fig.~\ref{mo}.  The differences between Obs 1, Obs 2 and Obs 3 can be modelled as differences in just the normalization and primary continuum photon index. Therefore $N$ and $\Gamma$ are denoted with underscores referring to the individual observations. The normalization is defined as usual in {\sc xspec}, photons keV$^{-1}$cm$^{-2}$s$^{-1}$ at 1~keV.}
\label{relxill}
\end{table}

To confirm that our model was appropriate for the soft band as well, we included the simultaneous {\em Swift} snapshots, and added Galactic absorption ($N_{\mathrm{H}}=5.8 \times 10^{20}$ cm$^{-2}$) and intrinsic absorption to the model.  This resulted in a $\chi^2/\mathrm{dof}=288/251=1.15$.  All parameters were the same within error, though the ionization parameter was slightly higher, log($\xi)=2 \pm 0.3$~erg cm s$^{-1}$.  The reflection fraction was constrained to be over 15.  The higher ionization is consistent with previous spectral studies of 1H0707-495 in the 0.3--10~keV band.  \citet{fabian12} and \citet{dauser12} found that the best fit required two reflectors with all parameters the same but with different ionization parameters.  The low-ionization reflection contributes most to the iron line, while the high-ionization dominates the soft excess. The double reflector is an empirical fit to the data, and is not motivated by accretion disc models.  In these papers, the double reflector was interpreted as a patchy accretion disc, where the density fluctuates throughout, but it could also be interpreted, for example, as an accretion disc that has a highly ionized surface layer, with a less ionized layer underneath.   

Given the small source height and large reflection fraction found in these observations, we know that the disc irradiation must be very strong. However, the ionization parameter is constrained to be log($\xi) \sim 0.3$ erg cm s$^{-1}$ in the {\em NuSTAR} band and up to log($\xi$)=2 erg cm s$^{-1}$ when including the soft excess, both of which are relatively low.  The ionization parameter $\xi = 4\pi F_{\mathrm{X}}/n_{\mathrm{e}}$, where $n_{\mathrm{e}}$ is the electron number density in the disc, and $F_{\mathrm{X}}$ is the irradiating X-ray flux from 1--1000 Ry.  If $F_{\mathrm{X}}$ is very high and the ionization parameter $\xi$ is low, this suggests that the density of the disc is very high in this object.  As an exercise, we compute a rough density of the disc given the luminosity and ionization parameter suggested from our fit to the 0.3--30~keV band.  The average 1--1000 Ry luminosity from our three observations is $1.2\times10^{43}$ erg s$^{-1}$.  Some of this luminosity is primary emission, and some is reflection.  We are interested in the amount of flux irradiating the disc. If we assume that the fraction of reflected flux ($R>15$) is equal to the fraction of the coronal emission that is irradiating the disc, then 15/16 of the total luminosity is irradiating the disc.  We convert this luminosity to a flux by dividing by $4\pi r^2$, where $r$ is the distance from the corona to the disc, which we find to be $\sim 2 r_{\mathrm{g}}$ or $6\times 10^{11}$~cm for a $2\times 10^{6}$ solar mass black hole \citep{bian03}.  Given the ionization parameter of $100$, we find that the electron number density, $n_e \sim 10^{17}$ cm$^{-3}$.  The {\sc xiller} and {\sc reflionx} grids assume reflection of a slab of $n_e \sim 10^{15}$ cm$^{-3}$. The implications of this density discrepancy may be important.

\begin{figure}
\includegraphics[width=\columnwidth]{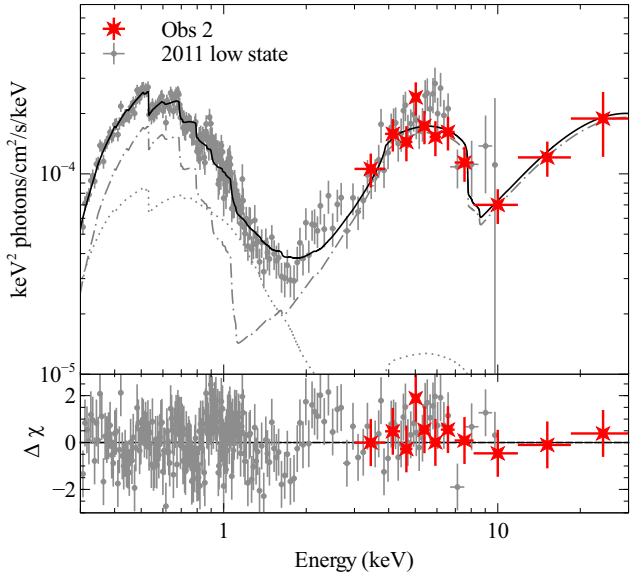}
\caption{The relativistic reflection model with two different ionization parameters fit to the low-flux observations: Obs~2 in the {\em NuSTAR} band (red), and the 2011 {\em XMM-Newton} ToO observation (grey).  The total model is the thick solid line. The low- and high-ionization reflectors are the dot-dash and dotted lines, respectively.  The model results in a reasonable fit, $\chi^2/{\mathrm{dof}}=403/355=1.14$.}
\label{mo_lo}
\end{figure}

\subsection{The low-flux spectrum}

Fig.~\ref{1h_nustar} shows that Obs~2 (red) was found in an extreme low flux state, very similar to the one triggered in 2011 with {\em XMM-Newton} (grey).  Unfortunately, there was no {\em Swift} data available at the time to confirm the soft band similarity between these observations.  However, we do have simultaneous UV coverage with the {\em Swift}-UVOT (for the {\em NuSTAR} observation), and with the Optical Monitor (for the {\em XMM-Newton} EPIC-pn observation).  The $V$-band magnitudes are remarkably similar for these two observations ($15.46 \pm 0.05$ with {\em Swift} and $15.47 \pm 0.01$ with {\em XMM-Newton}), which gives us confidence that we are looking at the source in the same flux level for the two observations.  Furthermore, the rms variability of the 3--10~keV band (the common band between the two observations) is the same within error. For a 80~ks length segment with 3000~s time bins, the {\em XMM-Newton} rms is $0.48 \pm 0.07$ and the {\em NuSTAR} rms is $0.65 \pm 0.28$, both of which are lower than the rms in high-flux observations.  The origin of the rapid variability in the low-flux observations is likely to include intrinsic variations in the X-ray source that cause correlated variations in the reflected emission.  The similarity of the rms in these two low-flux observations lends further support to fitting these two observations simultaneously.

We fit the {\em NuSTAR} and {\em XMM-Newton} low-state observations by extending our best fit {\sc relxilllp} model down to the {\em XMM-Newton} band from 0.3--10~keV.  Again, we include Galactic and intrinsic absorption into the model, and refit. 
The single absorbed relativistic reflection model to the {\em XMM-Newton} and {\em NuSTAR} low-flux observations gives a reasonable fit, $\chi^2/\mathrm{dof}=470/357=1.32$, with most of the deviations occurring between 1--2~keV.  
 The statistical fit is improved ($\chi^2/\mathrm{dof}=399/355=1.12$) if we include a second reflector that has all parameters tied except the ionization parameter and normalization, as in \citet{fabian12} and \citet{dauser13}.  Details are shown in Fig.~\ref{mo_lo} and Table~\ref{lo_table}.  The high-ionization reflector improves the fit at 1--2~keV.  The reduced $\chi^2$ is similarly improved if, instead of modelling a patchy disc, we add a separate multi-temperature black body component (i.e. {\sc diskbb} in {\sc XSPEC}) or a blackbody with temperature $kT=0.12$~keV, both of which fix the discrepancy at 1--2~keV.

\begin{table}
\centering
\begin{tabular}{l l}
\hline
{\bf Parameter} & {\bf Value}\\
\hline
$N_{\mathrm{H~(Gal)}}$ (cm$^{-2}$)$^{\alpha}$ & $5.8 \times 10^{20}$\\
$N_{\mathrm{H~(int)}}$ (cm$^{-2}$) & $<1 \times 10^{20}$\\
\hline
$h$ ($r_{\mathrm{g}}$) & $<1.4$\\
$a$ &$>0.988$ \\
$i$ ($^{\circ}$) & $65.0 \pm 14.0$\\
$r_{\mathrm{out}}$ $(r_{\mathrm{g}})^{\alpha}$& 400.0 \\
$\Gamma$ & $2.57 \pm 0.06$ \\
log($\xi_1$) (log(erg cm s$^{-1}$)) & $3.2 \pm 0.3$\\
log($\xi_2$) (log(erg cm s$^{-1}$)) & $1.2^{+0.03}_{-0.1}$\\
$A_{\mathrm{Fe}}$ & $>9.5$ \\
$E_{\mathrm{cut}}$ (keV)$^{\alpha}$ & 300.0  \\
$R$ & $>>10$\\
$N_1 \times 10^{-7}$ &$ 0.1 \pm 0.4$\\
$N_2 \times 10^{-7}$ &$ 10. \pm 0.4$\\
\hline
$\chi^{2}/\mathrm{dof}$ & 403/355 = 1.14 \\
\hline
\multicolumn{2}{l}{$^{\alpha}$ frozen}\\
\end{tabular}
\caption[Best-fit Spectral Parameters]{Best-fit spectral parameters of Galactic and intrinsically absorbed reflection model (with two different ionization parameters) to the low flux {\em XMM-Newton} and {\em NuSTAR} observations, corresponding to Fig.~\ref{mo_lo}.}
\label{lo_table}
\end{table}

With the double reflection model we constrain the low-flux state to have a lower source height than the fit to all the {\em NuSTAR} observations. Interestingly, when we include the {\em XMM-Newton} low-state spectrum the inclination becomes larger, and not well constrained.  One possibility for this  may be because the error bars above 7~keV are large in the {\em XMM-Newton} spectrum, and increasing the inclination helps to broaden the line in this very extreme spectrum.  It may also be that the inclination is increasing because the model is trying to fit the excess at 2--3~keV with the broad iron line.  It is possible that this 2--3~keV excess is not iron, but actually overabundant sulphur at $\sim 2.3$~keV. Unfortunately, we cannot account for overabundant sulphur with the current {\sc relxilllp} model.  The powerlaw flux is so low in this observation that we cannot constrain it with the current data. Therefore we cannot put strict constraints on the reflection fraction, though it is clearly very high. 

The self-consistent reflection fraction for a point source of $h=1.4$ is $R=8.2$. However, our source height is an upper limit, and therefore that self-consistent reflection fraction is a lower limit, which greatly increases as the source height decreases down to the event horizon (e.g. our true minimum in $\chi^{2}$-space is $h=1.17$, which corresponds to a reflection fraction of $R=22.3$).  If, instead of fitting the reflection fraction separately, we fit the reflection spectrum self-consistently for a point-source above the disc (fixReflFrac=1 in the {\sc relxilllp XSPEC} model), we obtain a fit with $\chi^2/\mathrm{dof}=536/358=1.49$. In this scenario, the high-ionization reflector is no longer required by the data, however this is still a statistically worse fit that the single reflector without a fixed reflection fraction ($\Delta \chi^{2}=63$ for one less degree of freedom).  The fit is much improved with an additional blackbody of temperature, $kT=0.12$~keV ($\chi^2/\mathrm{dof}=404/356=1.14$).  This tells us that there is some discrepancy between the data and the simple model of a static point-source above an infinite thin slab of density $n_{e}=10^{15}$~cm$^{-3}$.  These high-quality data are pushing the limits of our theoretical models, and therefore more work on the models is required, but is beyond the scope of this paper.

\section{Discussion}
\label{discuss}

\begin{figure*}
\includegraphics[width=0.85\textwidth]{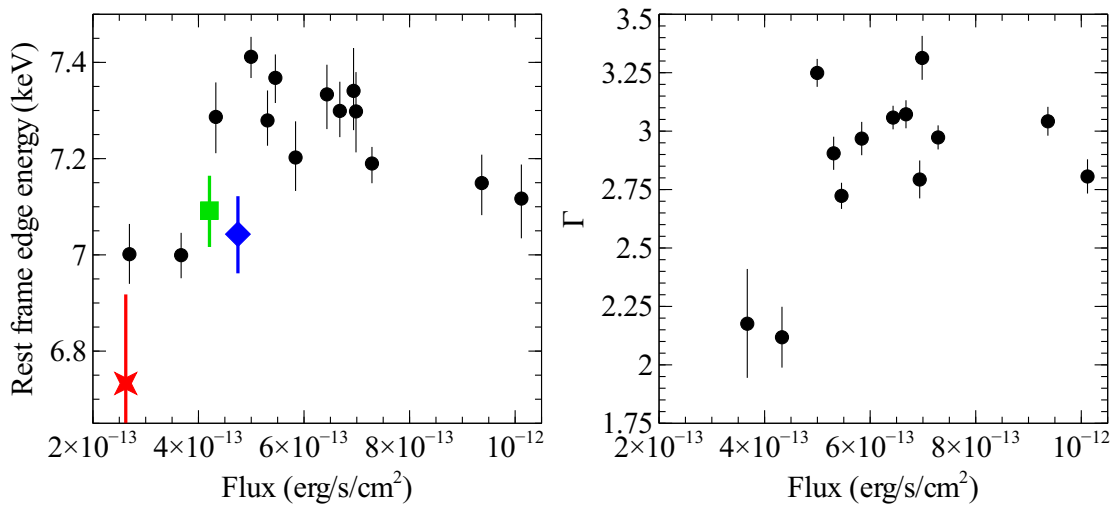}
\caption{{\em Left:} The rest-frame energy of the blue wing of the iron line versus the 3--10~keV flux.  The blue wing energy increases with flux up to $\sim 6 \times 10^{-13}$ erg/s/cm$^2$, above which the line energy scatters around 7.2~keV. {\em Right:} The powerlaw photon index from the `continuum' band (1.5--2.5~keV) versus the flux at 3--10~keV. We do not include our newest observations in this as the continuum cannot be well constrained from the short {\em Swift} snapshots.  We see a similar dependence on flux for the edge energy and the photon index, perhaps indicative of an underlying change in the corona with flux.}
\label{edge}
\end{figure*}


We have presented the analysis of the first {\em NuSTAR} observations of 1H0707-495.  The source went into an extreme low-flux state during these observation, but can still be observed up to 30~keV thanks to the unprecedented effective area  of {\em NuSTAR} in the hard X-ray band.  The data are well described by a relativistic reflection model, where the lamppost source height is constrained to be $\sim 2 r_{\mathrm{g}}$.  As in previous spectral studies \citep[e.g.][]{fabian09, dauser12}, the spin and iron abundance are very high, $a=0.97 \pm 0.03$ and  $A_{\mathrm{Fe}}>7.2$, respectively.  The reflection fraction is also very high in these observations, similar to what was found in \citet{fabian12} in a triggered {\em XMM-Newton} observation of 1H0707-495 in an extreme low state.  The high reflection fraction in low-flux observations can be explained by the strong gravitational light bending effects from a low coronal source height \citep{miniutti04}.  A high reflection fraction is also evidence for a high black hole spin \citep{dauser14,parker14}.

In this section, we discuss the apparent variable blue wing of the iron line, as seen by eye in Fig.~\ref{1h_nustar}.  We quantify this effect using all 15 archival {\em XMM-Newton} observations of 1H0707-495 in addition to these recent {\em NuSTAR} observations.  The photon index is observed to change with flux \citep{wilkins14}, and therefore it is important to get a good description of how the underlying continuum changes. To this end, we fit the 1.5--2.5~keV band with a power law for each observation.  We freeze the photon index and normalization of the underlying continuum, and then examine the 3--10~keV band (the iron line band).  We fit the excess above the continuum with an additional powerlaw with an edge.  In effect, we are fitting the red wing of the iron line as a power law, and the blue wing of the line as an edge.  This is a good diagnostic for measuring the energy of the blue wing.  All observations were well fit by this model. One 40~ks observations taken in 2007 (obsid: 0506200301) had too few counts to put meaningful constraints on the edge of the iron line, so for this observation, we fit it together with the following observation (obsid: 0506200201), which was contiguous with the first.  In Fig.~\ref{edge}-{\em left} we plot the best fit edge energy and the 3--10~keV flux for the 14 independent {\em XMM-Newton} observations in black, and our three new {\em NuSTAR} observations in green, red and blue.  We find that, especially for flux levels below $8 \times 10^{-13}$ erg~cm$^{-2}$~s$^{-1}$, the energy of the blue wing increases with increasing flux.  The edge energies of the two highest flux observations do not follow the increasing trend. A similar trend with flux is found in the continuum (1.5--2.5) keV powerlaw photon index (Fig.~\ref{edge}-{\em right}).  Similar to recent results by \citet{wilkins14}, the photon index increases with flux. However, again, the trend does not hold for the two highest flux observations.

\citet{gallo04} and \citet{fabian04} wrote back-to-back papers on the variable blue wing in 1H0707-495.  These papers were based on the first two {\em XMM-Newton} observations of the source.  They found that the spectra could be equally well explained by a changing partial covering absorber \citep{gallo04}, or by changes in the emissivity profile of the accretion disc with flux \citep{fabian04}.  If the data are described by the partial covering absorber, then the change in edge energy is best explained by a high speed outflow in the observation with the higher edge energy.  The absorber with variable edge energy from a Compton thick outflow was also used recently to describe the {\em XMM-Newton} observations \citep{mizumoto14}.  This interpretation would imply that the flux is increasing as the outflow velocity increases, which is possible if we assume that no new absorbers are generated and the covering fraction decreases with higher velocity outflow.  However, clumpy Compton thick absorption along our line of sight will result in a strong scattered component, unless the absorbers satisfy very restrictive geometric constraints \citep{reynolds09}. No clear narrow features are observed in the {\em XMM-Newton} EPIC-pn or RGS data, though, of course, this non-detection does not rule out the possibility of the specific geometry of multiple absorbers along our line that reabsorb all the scattered emission.  

The other interpretation explored to explain the shifting blue wing was changes in the coronal height in the relativistic reflection model. If in the low-flux state, the X-ray emitting source is compact and the source height is small, then only a small portion of the accretion disc will be irradiated due to strong gravitational light bending effects, and the observer will see mostly the gravitationally redshifted emission.  The flux increase can be explained as an increase in the coronal source height, which then causes larger radii to be irradiated.  This produces emission that is not gravitationally redshifted.  In addition to light bending effects, the structure of the disc itself could play a role in enhancing the gravitationally-redshifted emission at low flux levels. In a standard Shakura \& Sunyaev accretion disc near its Eddington limit, the scale height of the disc far from the innermost stable circular orbit can be as large as $5 r_{\mathrm{g}}$. If the coronal source height is less than $5~r_{\mathrm{g}}$, then, at medium inclinations, it would be eclipsed by the large scale height disc at larger radii.  The eclipse of the primary emission would cause the reflection fraction to increase, as we would still observe reflected emission from the far side of the disc (which is not eclipsed).  In this model, when the source height is higher, it will not be eclipsed and so the observed flux increases.  The large source height means that larger portions of the disc are irradiated, and the line centroid is clearly detected.  Both light bending and the disc eclipse would have the effect of steepening the emissivity profile with decreasing flux.  Similar affects have been found for the evolution of the emissivity break radius with flux in 1H0707-495 \citep{wilkins14}, and the steepening of the inner emissivity index with increasing reflection fraction in the `twin' to 1H0707-495, IRAS~13224-3809 \citep{chiang14}. 

The two highest flux observations do not agree with the trend of increasing edge energy with flux. One possible explanation for this is that the corona is more vertically extended in the high flux observations than in the ones below $8 \times 10^{-13}$ erg~cm$^{-2}$~s$^{-1}$. While more photons can escape the gravitational pull of the central black hole, there is still a significant portion of the continuum that is irradiating the innermost regions of the disc.  The fact that we also see this turnover at high fluxes for the photon index is also expected if the corona is more extended.  For example, in \citet{kara13a}, we showed that the energy spectrum of the high-frequency variability is harder than that of the low-frequency variability, suggesting that the corona emits a harder continuum at smaller radii. X-ray microlensing in quasars from \citet{chen11} also show that the harder X-ray emission originates from smaller radii than the soft emission.  The spectral hardening of the two highest flux observations would then suggest that some fraction of the coronal emission in these two observations is produced at small radii. We have not fit the archival {\em XMM-Newton} observations with an extended corona, and therefore we cannot quantify the extent. However \citet{wilkins13} has suggested through ray-tracing simulations of an extended corona, that 1H0707-495 is extended out to 20-30~$r_{\mathrm{g}}$, though most of the reflected emission comes from within $5~r_{\mathrm{g}}$. 

We also explore the effects of ionization on the broad line, but find that this is likely not the dominant effect.
As the ionization increases, there are K$\alpha$ transitions from more Fe ions, and we see emission from the H- and He- like ions (Fe {\sc xxv-xxvi}) up to $\sim 6.9$~keV.  This effect is only appreciable when the ionization parameter is high (above log($\xi$) $\sim 3$).  Our spectral fitting shows that the ionization increases with decreasing flux (consistent with stronger light bending in the low-flux state).  If ionization changes were driving the change in edge energy, then we would expect the opposite correlation with flux.  


Lastly, changes in disc inclination would also change the blue wing of the iron line, but this is not likely to be the cause.  If the disc were precessing, then the edge of the wing would decrease with increasing flux.  In the precessing disc model, as the inclination of the disc becomes smaller(more face-on), we see a higher flux because of the solid angle increase.  However, this would also decrease the Doppler boosting effects, and so the line would appear to be less broad (i.e. the energy of the blue wing of the line would decrease).  Furthermore, the second and third {\em NuSTAR} observations were taken days apart from each other, and the disc is unlikely to change on such short timescales.

The variable blue wing energy complicates our spectral modelling. If the changing emissivity profile affects the energy of the blue wing of the line, then we cannot use the blue wing as a clear indicator of the inclination of the accretion disc.  Usually, the blue wing energy is constant because it is determined by how much Doppler shifting is occurring. Therefore, it is likely that by fitting the {\em NuSTAR} data alone, we are inferring a lower disc inclination because it appears that there is less Doppler broadening.  

In order to account for this variable blue wing in our spectral modelling, we fit the {\em NuSTAR} observations with the same {\sc relxilllp} model, but this time freeze the inclination to 55.7 degrees. This is the inclination found in \citet{fabian09} from fitting a relativistic reflection model to 500~ks of {\em XMM-Newton} data when the 3--10~keV flux was $\sim 6 \times 10^{-13}$ erg~cm$^{-2}$~s$^{-1}$.  For this test, we take 55.7 degrees to be the `true' inclination, and use it to fit the recent {\em NuSTAR} spectra.  This model gives a good fit, $\chi^{2}/\mathrm{dof}$ = 265/230 = 1.15, and other parameters change slightly to account for this $>10$ degree difference in inclination.  The effect of increasing the inclination is generally to decrease the height of the source.  This, in return, increases the ionization and possibly the reflection fraction, which is to be expected as stronger gravitational light bending occurs when the source is lower.  When the inclination is fixed at 55.7 degrees, the height is more tightly constrained to be even lower, $h<1.5 r_{\mathrm{g}}$.  The ionization increases to log$(\xi)=1.6 \pm 0.7$~erg cm s$^{-1}$, and the reflection fraction is constrained to be slightly higher, $R>7.8$.  All other parameters remain largely the same. The spin is still high, $a=0.96 \pm 0.01$, and the iron abundance is still high $A_{\mathrm{Fe}}>8$.

We can track the changes in the energy of the blue wing of the iron line in 1H0707-495 because it has been observed for several years in many different flux states, but in addition to the impressive coverage, 1H0707-495 is special because it has a high iron abundance, which makes it easier to measure the iron features.  The question of the high iron abundance in 1H0707-495 has long been a criticism of the reflection modelling in this source.  Our new observations also require a high iron abundance.  In fact, if we ignore the {\em NuSTAR} data below 7.5 keV (i.e. below the iron line) and fit the high energy spectrum, we find that the fit requires an iron abundance greater than $2.9$ times solar abundance at the 68\% significance. While this is not extremely over-abundant or highly significant, it does suggest that the high iron abundance is not motivated by the iron~K edge alone, but also by the amount of photoelectric absorption in the 8--13~keV band.

\section{Conclusions}
\label{conclude}

We have presented the spectral analysis of the first {\em NuSTAR} observations 1H0707-495. Our main findings are:
\begin{itemize}
\item {\em NuSTAR} confirms a strong drop in flux at the iron~K band, and observes the Compton hump in this object for the first time. 
\item The {\em NuSTAR} observations were taken when the source was at a relatively low-flux.  The lowest flux observation (Obs~2), appears to be at an extreme low-flux state.
\item The three {\em NuSTAR} observations can be well fitted by a {\sc relxilllp} model with high spin and a high iron abundance.  The iron abundance derived from just fitting above 7.5~keV also suggests a super-solar iron abundance with 68\% significance.
\item The lowest flux observation, Obs~2, can be well fit with the extreme low-flux {\em XMM-Newton} observation with a model of relativistic reflection off a patchy (non-uniform density) disc. 
\item The energy of the blue wing appears to change with flux in this source.  The mostly likely explanation for this dependence on flux is that at low fluxes, the coronal source height is lower, and therefore we observe the more gravitationally redshifted emission at lower flux.  If we assume that the `true' inclination is the one found when the source was in a high-flux state and fit our {\em NuSTAR} spectrum again fixing the inclination to a higher value, then the source height for our new observations is constrained to be less than $h<1.5 r_{\mathrm{g}}$.
\end{itemize}

\section*{Acknowledgements}

EK thanks Javier Garcia and Thomas Dauser for interesting discussions on the {\sc relxilllp} modelling. EK the Gates Cambridge Scholarship. ACF thanks the Royal Society. EK, ACF, AM and GM acknowledge support from the European Union Seventh Framework Programme (FP7/2007-2013) under grant agreement n.312789, StrongGravity.  AM and GM acknowledge financial support from Italian Space Agency under grant ASI/INAF I/037/12/0-011/13.  This work was supported under NASA Contract No. NNG08FD60C, and made use of data from the {\em NuSTAR} mission, a project led by the California Institute of Technology, managed by the Jet Propulsion Laboratory, and funded by the National Aeronautics and Space Administration. We thank the {\em NuSTAR} Operations, Software and Calibration teams for support with the execution and analysis of these observations. This research has made use of the {\em NuSTAR} Data Analysis Software (NuSTARDAS) jointly developed by the ASI Science Data Center (ASDC, Italy) and the California Institute of Technology (USA).  This work made use of data supplied by the UK Swift Science Data Centre at the
University of Leicester.


\end{document}